\documentclass[aps,prl,twocolumn,superscriptaddress]{revtex4}

\usepackage{graphicx}

\usepackage{titlesec}

\titleformat{\section}[display]
        {\normalfont\small}{}{0pt}{\MakeUppercase}
             
\titlespacing*{\section}
{0pt}{12pt}{5pt}

\titlespacing*{\subsection}
{0pt}{5pt}{5pt}

\usepackage{amsfonts,amssymb,amsmath}

\usepackage{subfig}
\usepackage{color}

\newcommand{\rmd}{{\, \mathrm d}}
\newcommand{\rme}{{\, \mathrm e}}
\newcommand{\rmi }{{\mathrm i}}

\newcounter{subequation}
        {\addtocounter{equation}{-1}%
        \stepcounter{subequation}%
        \begin{equation}}%
        {\end{equation}%
}

\newcommand{\rmdS}{\circ \rmd}

\newcommand{\esq}{{\mathbb{E}}}
\newcommand{\ninf}{{N\rightarrow\infty}}

\begin{document}
\title{Brownian regime of finite-$N$ corrections \\ to particle motion in the XY hamiltonian mean field model}

\author{Bruno V. Ribeiro}
\email{bruno.ribeiro@ifg.edu.br}
\affiliation{Instituto Federal de Educa\c{c}\~ao, Ci\^encia e Tecnologia de Goi\'as - C\^ampus Valpara\'{\i}so,
		       	 BR 040, Km 6, \'Area 8, Gleba E, \\ 
			 Fazenda Saia Velha, Parque Esplanada IV - Valpara\'{\i}so de Goi\'as - GO, Brasil}

\author{Marco A. Amato}
\affiliation{Instituto de F\'isica, Universidade de Bras\'{\i}lia, 
 				CP: 04455, 70919-970 - Bras\'{\i}lia - DF, Brasil\\
 			International Center for Condensed Matter Physics, 
Universidade de Bras\'{\i}lia, DF - Brasil}

\author{Yves Elskens}
\affiliation{Equipe turbulence plasma, case 322, 
		      	PIIM, UMR 7345 CNRS,\\
		       	Aix-Marseille Universit\'e,
		       	campus Saint-J\'er\^ome, FR-13397 Marseille cedex 13, France}

\begin{abstract}
     We study the dynamics of the $N$-particle system evolving in the XY hamiltonian mean field (HMF) model
     for a repulsive potential, when no phase transition occurs.
     Starting from a homogeneous distribution, particles evolve in a mean field created by the interaction with all others. 
     This interaction does not change the homogeneous state of the system, 
     and particle motion is approximately ballistic with small corrections.
      For initial particle data approaching a waterbag,
      it is explicitly proved that corrections to the ballistic velocities are in the form of independent brownian noises 
      over a time scale diverging not slower than $N^{2/5}$ as $N \to \infty$,
      which proves the propagation of molecular chaos. Molecular dynamics
      simulations of the XY-HMF model confirm our analytical findings.
\end{abstract}

\maketitle

\section{Introduction}
\label{SecIntro}

The dynamics of long-ranged interacting systems are a topic of active investigation due to their intriguing properties \cite{CaDaRu09}. As particles interact with every other ones in these systems, collective behaviour is prone to dominate over collisional processes during long times. This interplay between collisional relaxation and collective behaviour is responsible for the richness of phenomena in long-ranged systems, as well as for their unusual relaxation towards equilibrium. For example, starting from an initial nonequilibrium configuration, these systems rapidly evolve to a quasistationary state (QSS) where they are trapped for long times \cite{AnRu95,BeTePaLe12}. These times scale with the number of particles in the system and are followed by a proper relaxation towards thermodynamical equilibrium. For the complexity of their dynamics, these systems have raised interest in various fields such as plasma physics, astrophysics, statistical mechanics and applied mathematics (see \cite{CaDaRu09,DaRuAWi02,DaRuCu10,LePa14} for reviews).

Dynamical properties, including the intricate collective behaviour of constituents, of these systems may be unveiled through one-dimensional finite $N$ models of mean-field type \cite{CaDaRu09}. Despite their simplicity, these one-dimensional models present many of the rich dynamical properties of long-ranged systems, such as the emergence of QSS, where the time average of macroscopic quantities differs from their ensemble averages in statistical equilibrium. 
One such system is the XY hamiltonian mean field model (HMF), in which $N$ identical particles interact via a infinite-ranged potential and have their motion confined to a unit circle. This system evolves in phase-space $S^N_{2\pi} \times \mathbb{R}^N$ according to the hamiltonian \cite{AnRu95}
\begin{equation}
	H = \sum_{i=1}^N \frac{p_i^2}{2} - \frac{1}{2N} \sum_{i,j=1}^N V \cos(q_j-q_i),
\label{ham}
\end{equation}
where the $i$-th particle has unit mass, position $q_i$ and momentum $p_i$, $S_{2\pi}=\mathbb{R}/{2\pi}$ and the constant $V$ defines the nature of the interaction. In this system, each particle interacts with all other ones through a force field that is, at each instant, the sum of the individual fields produced by all particles. The interaction term in the hamiltonian is equivalent to the interaction term in the XY Heisenberg model (or ``rotator model'') in the mean-field approach. The model with positive interactions ($V>0$) corresponds to the ferromagnetic case, while the negative interaction ($V<0$) corresponds to the antiferromagnetic case.

One can understand the complexity of the dynamics of this seemingly simple system by analysing the equations of motion
\begin{eqnarray}
	\dot{q}_i &=& p_i \label{eomhmfq} \\
	\dot{p}_i &=& -\frac{V}{N} \sum_{j=1}^N \sin(q_i - q_j),
\label{eomhmfp}
\end{eqnarray}
which correspond to a set of $N$ fully coupled \textit{pendula}.

Introducing the mean-field quantity (as a reference to the Heisenberg model we call it ``magnetization")\
\begin{equation}
	\mathbf{M} = \frac{1}{N}\sum_{i=1}^N \rme^{\rmi q_i} \equiv M \rme^{\rmi \varphi},
\label{mag}
\end{equation}
we write the equations of motion (e.o.m.)
\begin{equation}
	\ddot{q}_i = - V M \sin(q_i - \varphi).
\label{eom}
\end{equation}

Therefore, the motion of each particle is determined by a self-consistent interaction with a mean-field $\mathbf{M}$, which depends on time implicitly through the instantaneous values of particle positions. The average energy per particle is given by
\begin{equation}
	U = \frac{H}{N} = \frac{\left\langle p^2 \right\rangle }{2} + \frac{V}{2}(1 - M^2),
\label{enmedppart}
\end{equation}
where $\left\langle \cdot \right\rangle $ represents average over all particles. From this last expression, we can use $\mathbf{M}$ as an order parameter to characterize the phases of the system. For the ferromagnetic case ($V=1$), the ground state corresponds to a state in which all particles are clustered, giving a high value for $M$; in the antiferromagnetic state ($V=-1$), the ground state corresponds to a uniform distribution of particles in position space, resulting in a zero value for $M$.

Equilibrium statistical mechanics calculations in the canonical ensemble predict a phase transition in the ferromagnetic case between a clustered phase, with non-zero values for $M$, and a homogeneous phase, in which $M$ vanishes  \cite{AnRu95,BaDaRu01}. 
For the antiferromagnetic case, the only equilibrium solution is the homogeneous state in which $M=0$. 
Though microcanonical calculations are more difficult,
it has been proved that both ensembles are equivalent for (\ref{ham}) \cite{CaDaRu09},
and it has been numerically checked that, for the HMF model, ensembles are equivalent 
for high enough energies\footnote{The only exception to this case is the bicluster formation in the antiferromagnetic case that happens at very low energies \cite{BaDaRu01}.} 
\cite{DaLaRaRuTo02}.
Since molecular dynamics (MD) for the system (\ref{eomhmfq})-(\ref{eomhmfp}), as we consider below, involve no heat bath,
the relevant Gibbs ensemble is the microcanonical one.

In the $\ninf$ limit, the description of the system is rigorously given by the Vlasov equation \cite{Sp91}. 
This equation governs the evolution of the single particle distribution function $f(q,p,t$). 
For long-ranged interacting systems, the QSSs that follow the initial nonequilibrium configuration represent stable steady states of the underlying Vlasov equation.

However, for finite systems, in the antiferromagnetic case, MD simulations show that starting from an initial homogeneous configuration, the value of $M(t)$ fluctuates around small values due to density fluctuations of particles in the circle $S_{2\pi}$. Fluctuations as this are, usually, assumed to have a diffusive nature based on empirical data from molecular dynamics simulations \cite{EtFi11}. 

In this note, we work out the dynamical details that lead to the fluctuations in the mean-field quantity $\mathbf{M}$, showing the nature of the underlying process driving the motion of the particles. 
Thanks to the simplicity of the XY-HMF model, we provide a direct, short proof that these fluctuations 
generate independent diffusion of particles on long time-scales.



\section{Ballistic motion and corrections}

Because there is no cluster formation in the stable state of the repulsive case, 
a first approximation to the particle motion is the ballistic motion
\begin{equation}
	q_j(t) = q_{j0} + v_{j0}t + Q_j,
\label{ballapp}
\end{equation}
where $v_{j0}$ is a constant velocity associated with particle $j$ and $Q_j$ is a small correction to ballistic motion ($Q_j \ll 2 \pi$), which we will neglect for the moment.
So, we rewrite
\begin{equation}
	M \rme^{\rmi \varphi} \simeq M^{(0)} \rme^{\rmi \varphi^{(0)}} := 
	\frac{1}{N}\sum_{j=1}^N \rme^{\rmi(q_{j0} + v_{j0}t)} 
\label{repmag}
\end{equation}
and
\begin{equation}
	\ddot{q}_j \simeq M^{(0)} \sin(q_j - \varphi^{(0)}).
\label{repeom}
\end{equation}

To study the nature of the evolution of the system, our goal is to write the evolution equation \eqref{repeom} as a differential equation driven by a known (possibly wild, rough or stochastic) process. 
As we expect a small noise, its effects will show up for large times, so we rescale time to $t'=t/N''$ 
with some large $N''$ ; simultaneously, 
assume we can write the number of particles as the product of two integers $N = N' \cdot N''$ and consider
\begin{equation}
	\mathbf{M} \, \rmd t 
	= \frac{1}{N} \sum_{j=1}^N \rme^{\rmi(q_{j0} + v_{j0}t)} \, \rmd t 
%
	= N'^{- \alpha} \sqrt{2\pi} \, \rmd W_{t'}^N ,
\label{resmag}
\end{equation}
where we introduce the complex-valued process
\begin{equation}
	\rmd W_{t'}^N 
	= \frac{1}{\sqrt{2\pi} N'^{1-\alpha}} \, \sum_{j=1}^N \rme^{\rmi q_{j0}}\rme^{\rmi N'' v_{j0}t'} \, \rmd t'
\label{1W}
\end{equation}
with an exponent $0 < \alpha < 1$ to be fixed later. 

With this definition, we write the e.o.m.\ to dominant order as
\begin{eqnarray}
	\rmd q_i &=& N'' p_i \, \rmd t', 
	\\
	\rmd p_i &=& \sqrt{2\pi} N'^{- \alpha} \, [\sin(q_i) \, \rmd \Re(W_{t'}^N) - \cos(q_i) \, \rmd \Im(W_{t'}^N)]. 
	\nonumber \\ & &
\label{1eom}
\end{eqnarray}
To keep the evolution of rescaled quantities independent of system size, 
we rescale momentum to $p'_i = N'' p_i$ and 
we balance scalings $N'' = N'^\alpha$ to get
\begin{eqnarray}
	\rmd q_i &=& p'_i \, \rmd t', 
	\label{2eoma} 
	\\
	\rmd p'_i &=& \sqrt{2\pi} \, [\sin(q_i) \, \rmd \Re(W_{t'}^N) - \cos(q_i) \, \rmd \Im(W_{t'}^N)].
\label{2eomb}
\end{eqnarray}

Note that the interaction potential in \eqref{ham} is well behaved, so that microscopic motion is smooth for finite $N$, thus (\ref{2eoma})-(\ref{2eomb}) involve ordinary calculus with differentials $\rmd \Re(W_{t'}^N)$ and $\rmd \Im(W_{t'}^N)$.

Consistently, we set
\begin{equation}
	v'_{j0} = N''v_{j0},
\label{beams}
\end{equation}
so that
\begin{equation}
	\rmd W_{t'}^N 
	= \frac{N''^{1/2}}{N'^{1-\alpha}} \rmd W_{N''}^{N'}(t') 
\label{2Wa}
\end{equation}
with the rescaled process defined by 
\begin{equation}
     \rmd W_{N''}^{N'}(t')
	= 
	\frac{1}{\sqrt{2\pi N''}} \sum_{j=1}^{N'N''} \rme^{\rmi q_{j0}}\rme^{\rmi v'_{j0}t'}\rmd t'.
\label{2Wb}
\end{equation}

\begin{figure}
\centering
\subfloat[][]{%
\label{fig:ex3-a}%
\includegraphics[width=6cm]{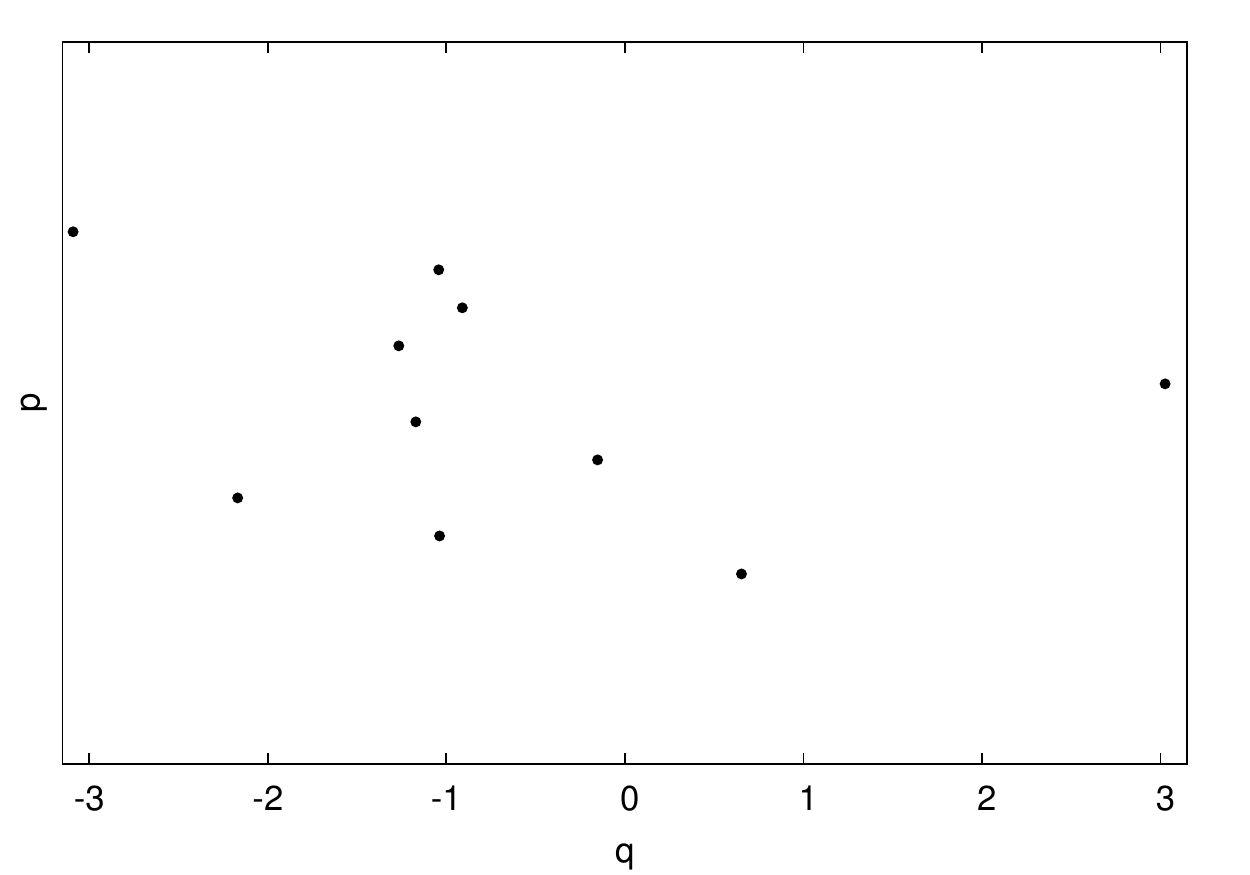}}%
\hspace{8pt}%
\subfloat[][]{%
\label{fig:ex3-b}%
\includegraphics[width=6cm]{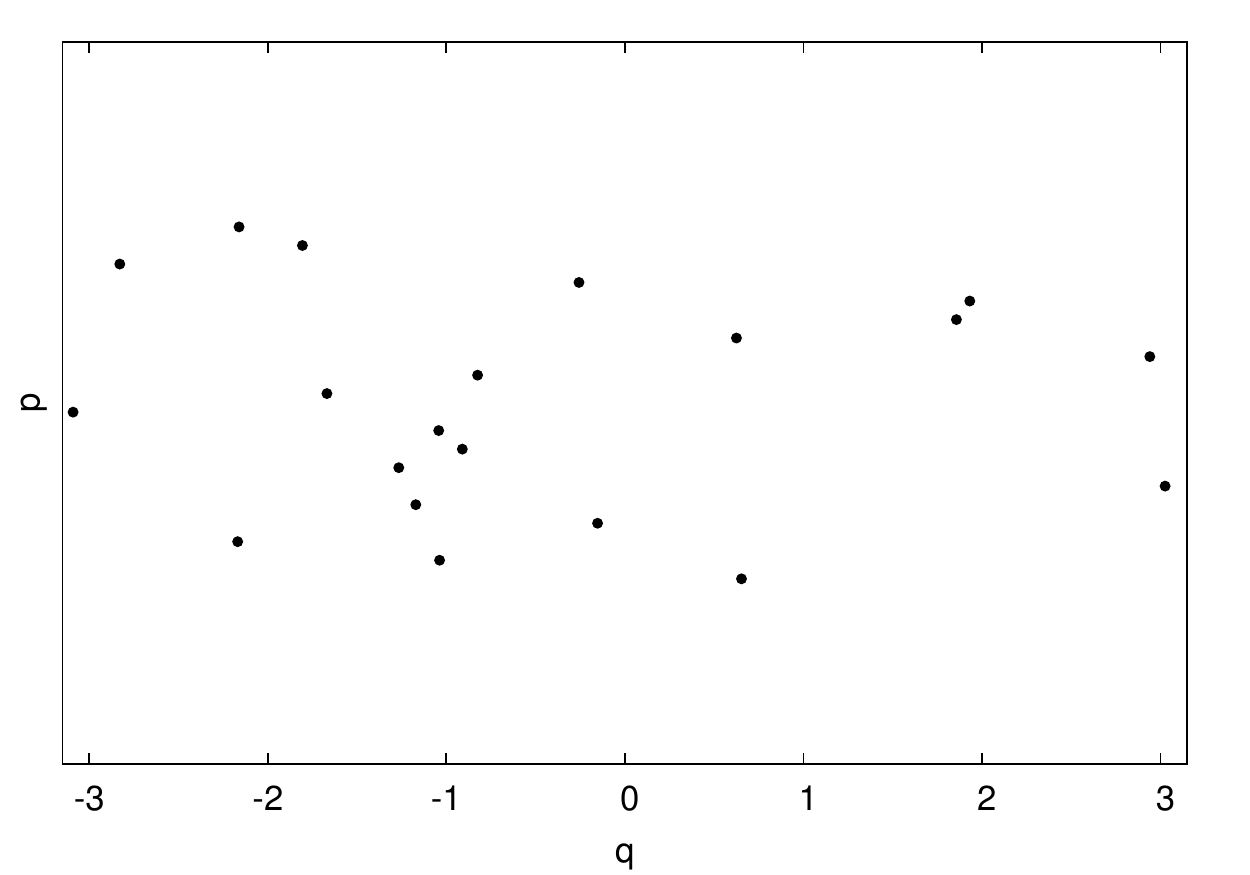}}\\
\subfloat[][]{%
\label{fig:ex3-c}%
\includegraphics[width=6cm]{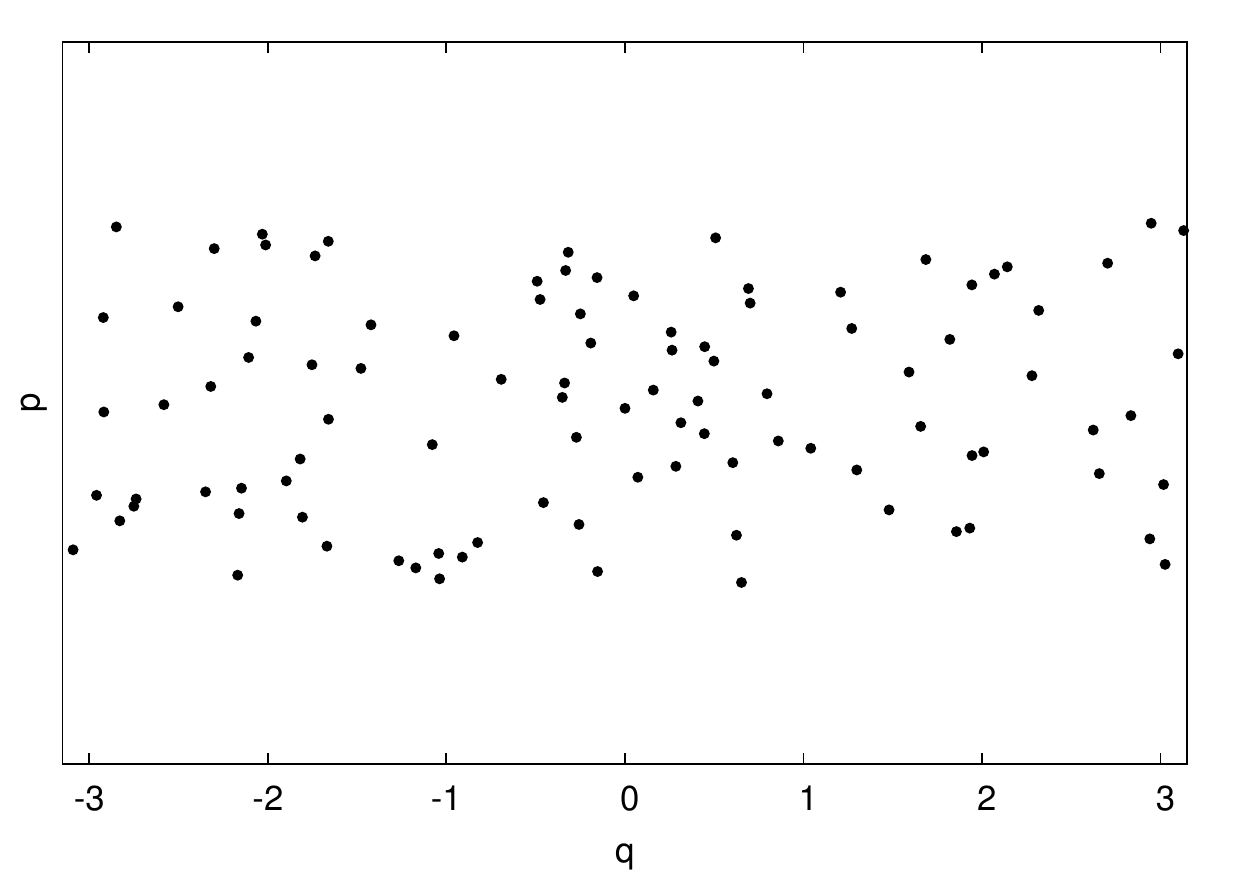}}%
\hspace{8pt}%
\subfloat[][]{%
\label{fig:ex3-d}%
\includegraphics[width=6cm]{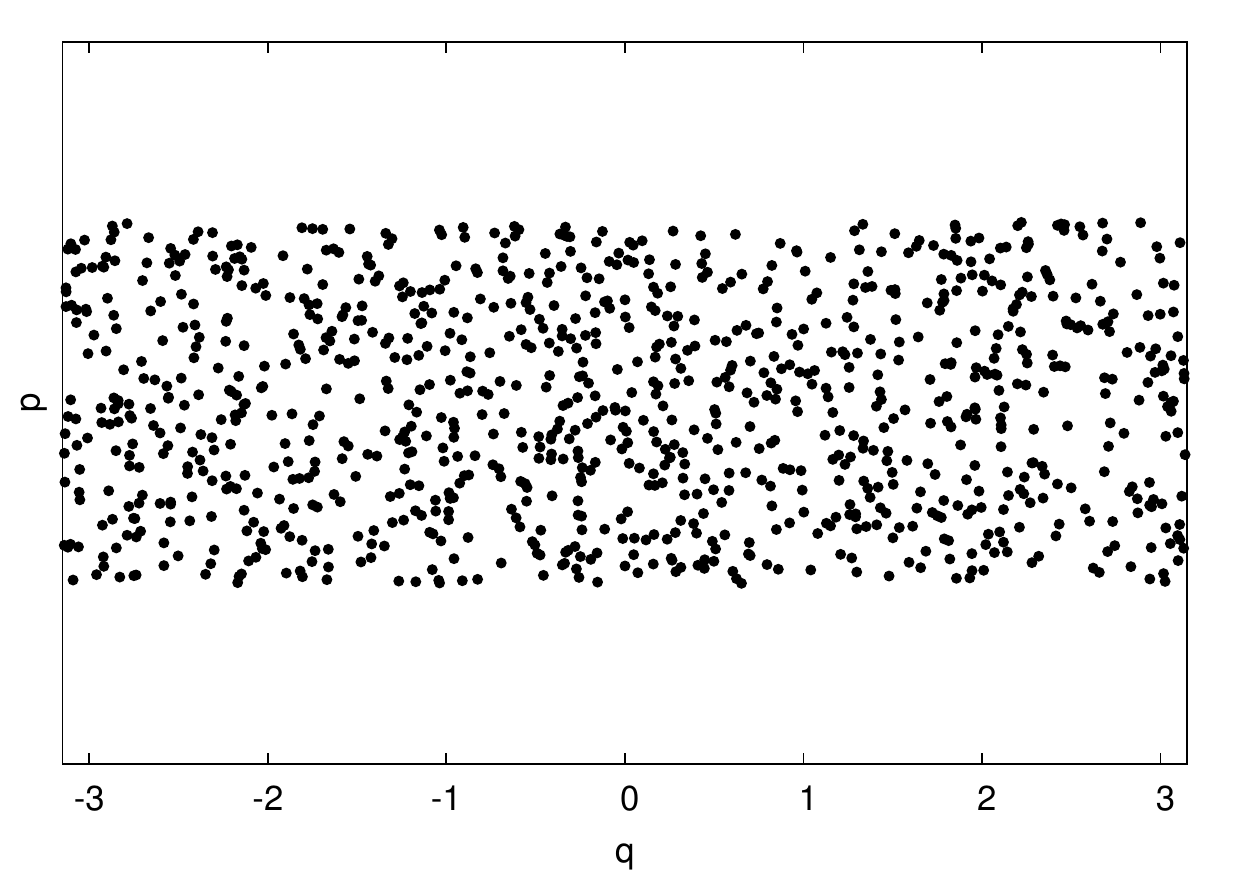}}%
\caption[Condições iniciais.]{\footnotesize{Initial distribution of particles in monokinetic beams in the $(q,p)$ space. Distribution of:
\subref{fig:ex3-a} 10 particles, \subref{fig:ex3-b} 20 particles,
\subref{fig:ex3-c} 100 particles and \subref{fig:ex3-d} 1000 particles. We can see how the distribution approaches a ``waterbag"\ on increasing $N$.}}
\label{inibeam}%
\end{figure}

The particles are taken to be, initially, distributed independently, uniformly in position. In velocity space, it will be useful to set particles to be distributed, initially, in equally spaced beams containing one particle each. 
This guarantees that, in the $N \to \infty$ limit, the particle distribution converges to a ``waterbag''  \footnote{The ``waterbag"\ distribution 
is that in which particles are distributed uniformly in a region of $(q,p)$ space. 
For a rectangle $[-q_0,q_0] \times [-p_0,p_0]$, this one-particle distribution function is
$$	f_{\rm{WB}} = \frac{1}{4q_0 p_0}\Theta(q_0 - |q|)\Theta(p_0 - |p|), $$
where $\Theta$ is the Heaviside function.}
over a band $S_{2\pi} \times [-p_0, p_0]$ (see figure \ref{inibeam}). 
In the limit $N \to \infty$, this distribution is a stationary solution to the Vlasov equation, 
and a standard initial data for numerical simulations \cite{RoAmFi11,RoAmFi12}. 

To implement our ``beam"\ model, we write
\begin{equation}
	v'_{j0} = \frac{j - 1/2}{N''} - \frac{N'}{2}
\label{vbeam}
\end{equation}
and reshape the $N$-dimensional vectors $q_{j0}$ and $v'_{j0}$ into $N' \times N''$ matrices $q^0_{m,n}$ and $v'^{\, 0}_{m,n}$ with $j = n + (m-1) N''$ to get the proper limit in the next section. 
In this spirit, the new representation for initial velocity is
\begin{equation}
	v'^{\, 0}_{m,n} = (m-1) + \frac{n - 1/2}{N''} - \frac{N'}{2},
\label{vmat}
\end{equation}
where $m = 1,...,N'$ and $n = 1,...,N''$, while the initial positions are simply $q^0_{mn}$ and remain uniformly distributed. With this new formulation, (\ref{2Wb}) reads
\begin{equation}
	\rmd W_{N''}^{N'}(t') 
	= 
	\frac{1}{\sqrt{2\pi N''}} \, \sum_{n=1}^{N''} \sum_{m=1}^{N'} 
	            \rme^{\rmi q^0_{mn}} \rme^{\rmi[(m-1) + (2n-1)/(2 N'') - N'/2]t'} \rmd t' 
\label{3W}
\end{equation}
where one recognizes in $N''$ a number of samples for taking a central limit theorem 
and in $N'$ a bandwidth scale which contributes higher frequencies and allows for a non-smooth limit process $W$.

\section{The $N \rightarrow \infty$ limit}


According to \cite{El12}, in the limit $N',N'' \to \infty$, 
the process $W_{N''}^{N'}(t')$ approaches a Wiener process ($W_{t'}$) in $\mathbb{C}$. 
Thus, calculus with differentials $\rmd \Re(W_{t'}^N)$ and $\rmd \Im(W_{t'}^N)$ becomes Stratonovich's \cite{WoZa65}, denoted with $\rmdS$,
\begin{eqnarray}
	\rmd q_i &=& p'_i \rmd t', \\
	\rmd p'_i &=& \sqrt{2\pi} \, \frac{N''^{1/2}}{N'^{1-\alpha}} \, 
	                       [\sin(q_i)\rmdS \Re(W_{t'}) - \cos(q_i)\rmdS \Im(W_{t'})] ,
	\nonumber \\ & &
\label{eomninf}
\end{eqnarray}
and we select the scaling $N''^{1/2} = N'^{1-\alpha}$. 
Along with $N'' = N'^\alpha = N / N'$, this yields $\alpha = 2/3$,
$N' = N^{3/5}$ and $N'' = N^{2/5}$. 
By Proposition 4.2 of \cite{El12}, these equations yield
\begin{eqnarray}
	q_i(t') &=& q_{i0} + p'_{i0} t' + \sqrt{2\pi} \, \int_0^{t'} B_i(s)\rmd s , \\
	p'_i(t') &=& p'_{i0} + \sqrt{2\pi} \, B_i(t'),
\label{1dyn}
\end{eqnarray}
where $B_i$ is one realization of the standard one-dimensional brownian motion.

Computing the variations of (\ref{1dyn}), 
\begin{eqnarray}
	\delta q_i(t') & \sim & t'\delta p_i' \sim {\mathrm{O}}\left(t'^{3/2}\right) , \\
	\delta p'_i(t') & \sim & \sqrt{2\pi} \, B(t') \sim {\mathrm{O}} \left(t'^{1/2} \right).
\label{var}
\end{eqnarray}

So, as a correction to ballistic motion, we get
\begin{equation}
	q_j(t) = q_{j0} + v_{j0}t + \delta q_i(t') \sim q_{j0} + v_{j0}t + {\mathrm{O}}\left( \frac{t}{N''} \right)^{3/2}.
\label{1corrball}
\end{equation}
The ballistic approximation is valid, then, at least for
\begin{equation}
	\frac{t}{N''} \ll 1,
\label{1val}
\end{equation}
which is guaranteed for any finite time in the $N',N'' \to \infty$ limit. 
For finite $N$, the approach breaks down when $t \sim N'' \sim N^{2/5}$. 

Moreover, any fixed number of particles are independent in the limit, 
viz.\ ``molecular chaos propagates'' \cite{El12,Kac56,Kac59}.

\goodbreak
\section{Numerical results}

To test our analytical findings, we simulate the evolution of the system ruled by the hamiltonian \eqref{ham} 
and observe the fluctuations in particle velocity. 
We use a molecular dynamics code with a fourth order symplectic numerical integrator \cite{Yo90}.

\begin{figure}
\centering
\subfloat[][]{%
\label{inicond}%
\includegraphics[width=0.45\textwidth]{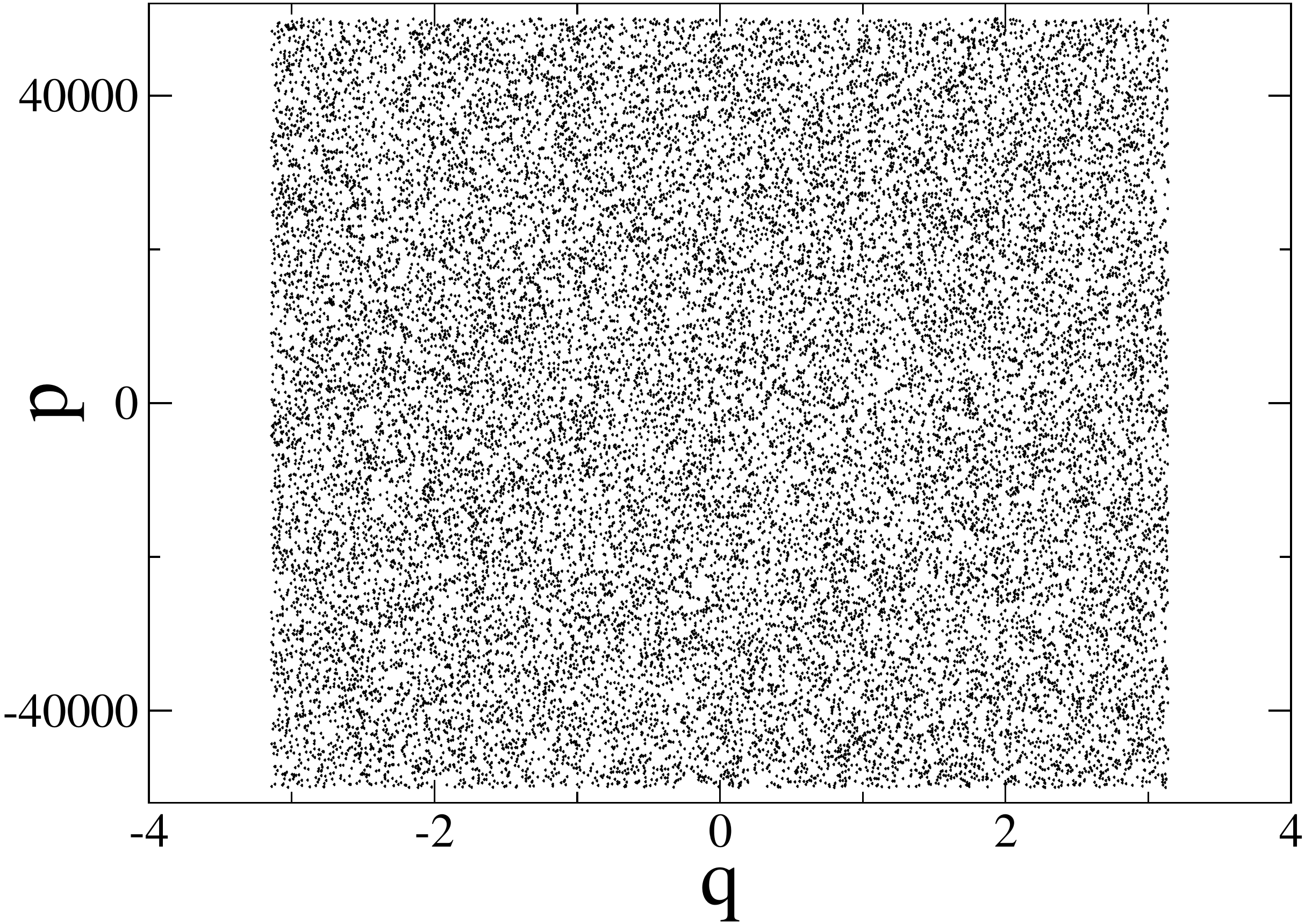}}%
\hspace{2pt}%
\subfloat[][]{%
\label{fincond}%
\includegraphics[width=0.45\textwidth]{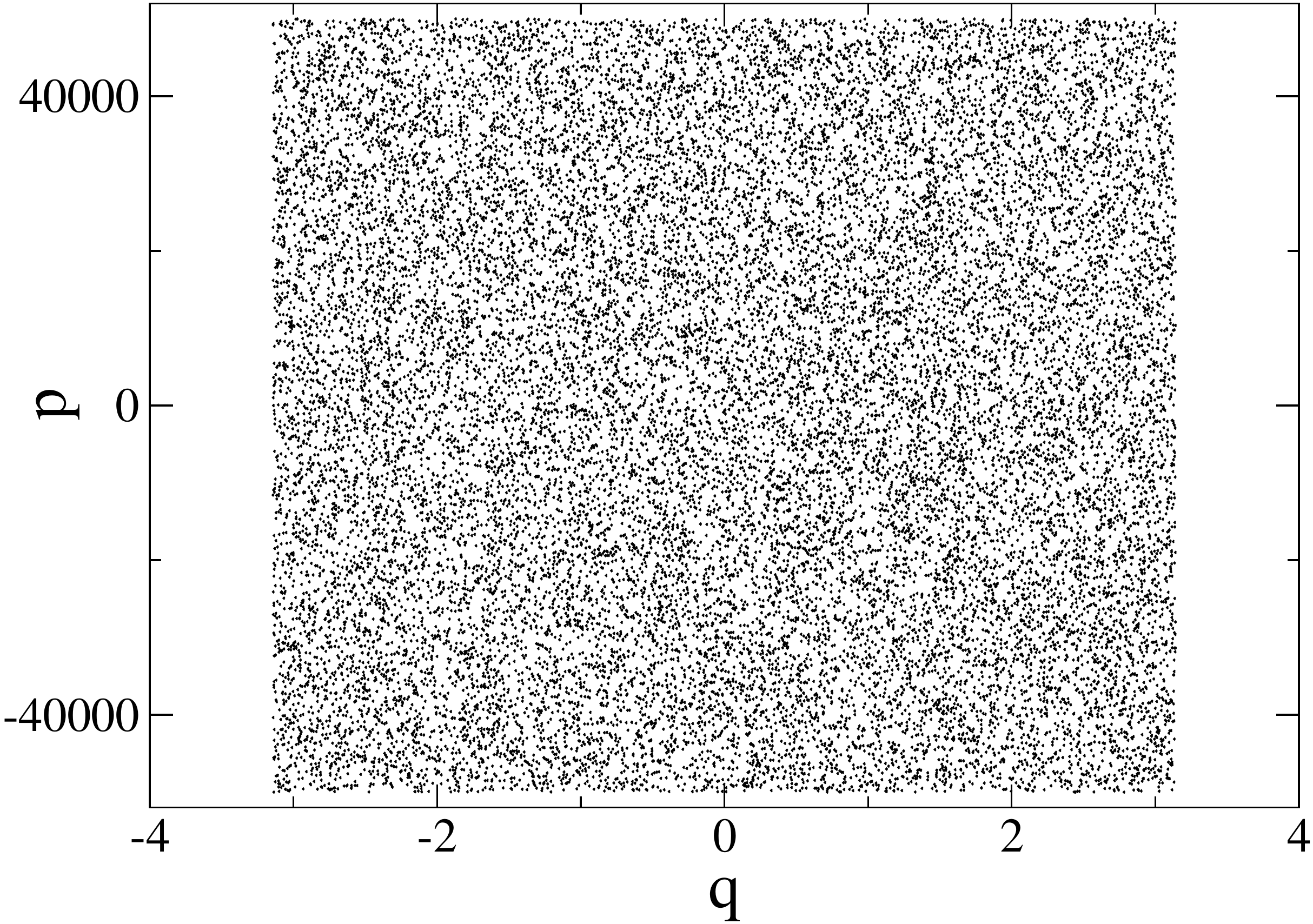}}
\caption[Ev fase]{\footnotesize{$(q,p)$-space portraits:
\subref{inicond}  initial state, \subref{fincond} state at $t=200$. Note that the system remains in a homogeneous state. System size $N=10^5$. }}
\label{phport}%
\end{figure}

We note that condition \eqref{vbeam} ensures that the (conserved) total momentum is exactly zero. 
The $(q,p)$-space portrait for this initial configuration (at $t=0$), 
with a uniform distribution of positions in the interval $[-\pi, \pi]$, is presented, 
along with the $(q,p)$-space portrait for a later time ($t=200$), in figure \ref{phport}.

\begin{figure}[hbtp]
\centering
\includegraphics[scale=0.3]{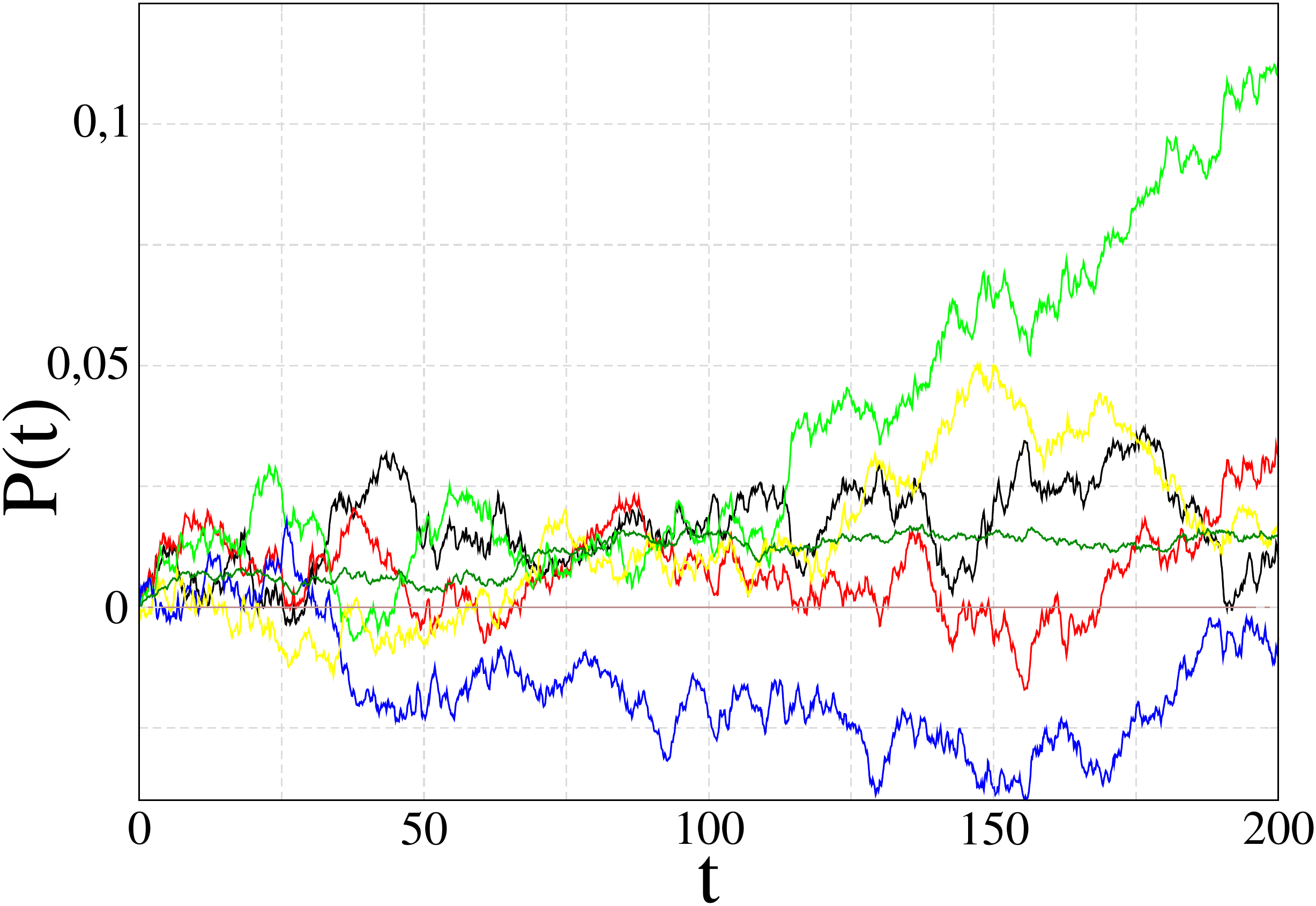}
\caption{Evolution of velocity for 6 different particles. Initial conditions given by \eqref{vbeam}.}
\label{paths}
\end{figure}

As a first illustration of the motion of particles in velocity space, we simulate the system with $N = 10^6$ particles, randomly choose 6 particles and plot each of the respective $P_j(t)= p_j(t) - p_{j0}$ with a timestep of the numerical integrator $\delta t = 0.1$. Results are shown in figure \ref{paths}.

\begin{figure}[hbtp]
\centering
\includegraphics[scale=0.3]{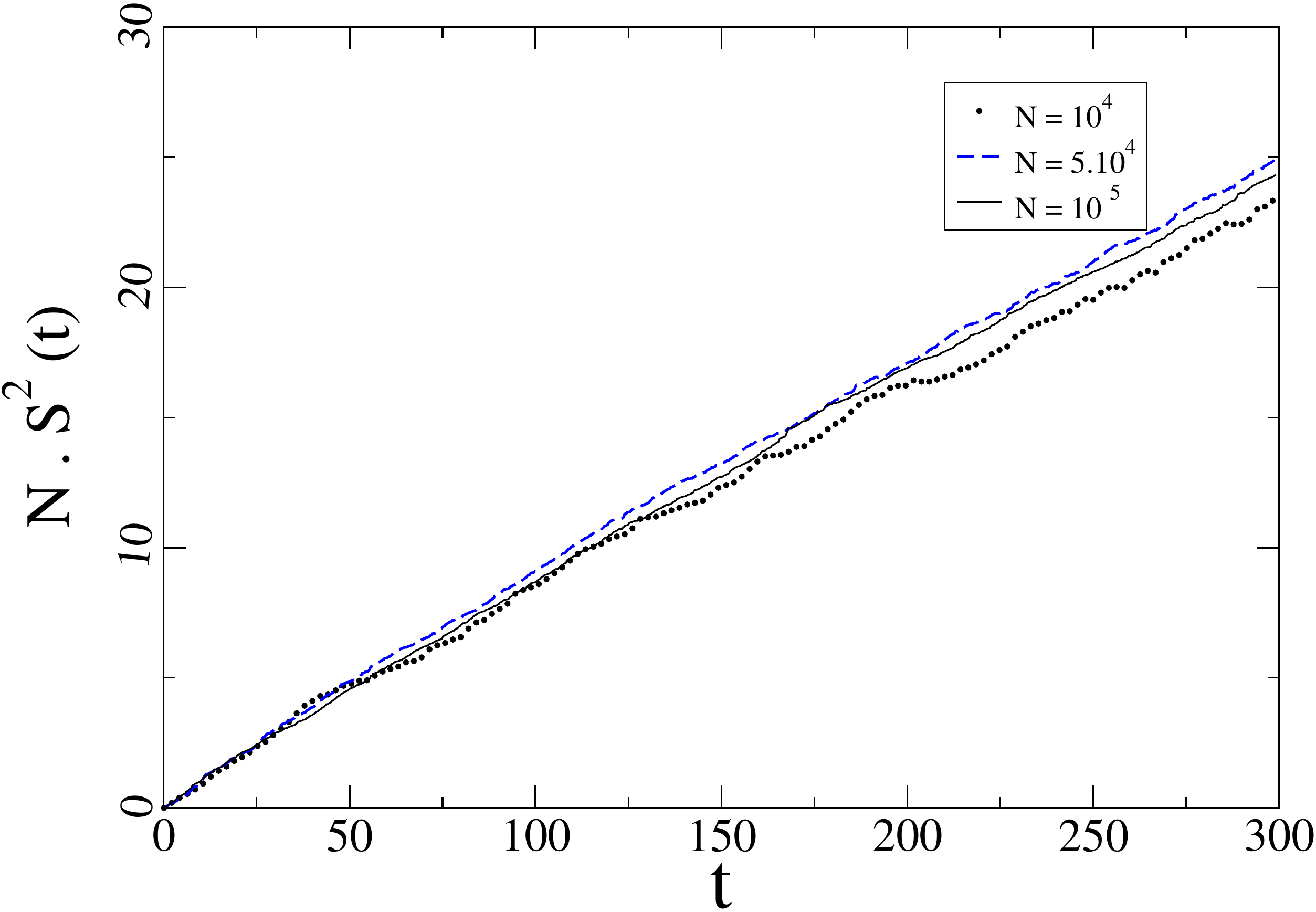}
\caption{Evolution of momentum deviation average square $S^2(t)$. 
System sizes $N = 10^4$, $N = 5 \cdot 10^4$ and $N = 10^5$, sample size $N_{\mathrm{s}} = 0.1N$, 
total simulation time $t = 300$, time step $\delta t = 0.1$. }
\label{squared}
\end{figure}

It is known that a Wiener process $W(t)$ is a Gaussian process with independent increments characterized by
\begin{equation}
	W(0) = 0, \quad \esq(W(t)) = 0, \quad {\rm{Var}}(W(t)) = t,
\label{wienercarac}
\end{equation}
where $\esq$ is the expectation operator and $\rm{Var}$ the variance. Therefore, to test whether our $P_j$ is indeed a brownian motion, we define the operator $\left\langle \bullet \right\rangle $ by its action on a process $R_j$ as
\begin{equation}
	\left\langle R_j \right\rangle  = \frac{1}{N_{\mathrm{s}}} \sum_j R_j ,
\label{defavg}
\end{equation}
where $N_{\mathrm{s}}$ is the sample size. With this, introduce the quantity
\begin{equation}
	S(t) = \sqrt{ \left\langle P_j^2(t) \right\rangle },
\label{st}
\end{equation}
and verify whether $S^2(t)$ grows linearly with time $t$, in agreement with \eqref{wienercarac}. 
Results are presented in figure \ref{squared}.

We get a linear behaviour for $S^2(t)$ in various system sizes. 
Note that the value of $S(t)$ scales as $N^{-1/2}$, as the evolutions of $N\cdot S^2(t)$ coincide for the simulated systems.

\begin{figure}[hbtp]
\centering
\includegraphics[scale=0.35]{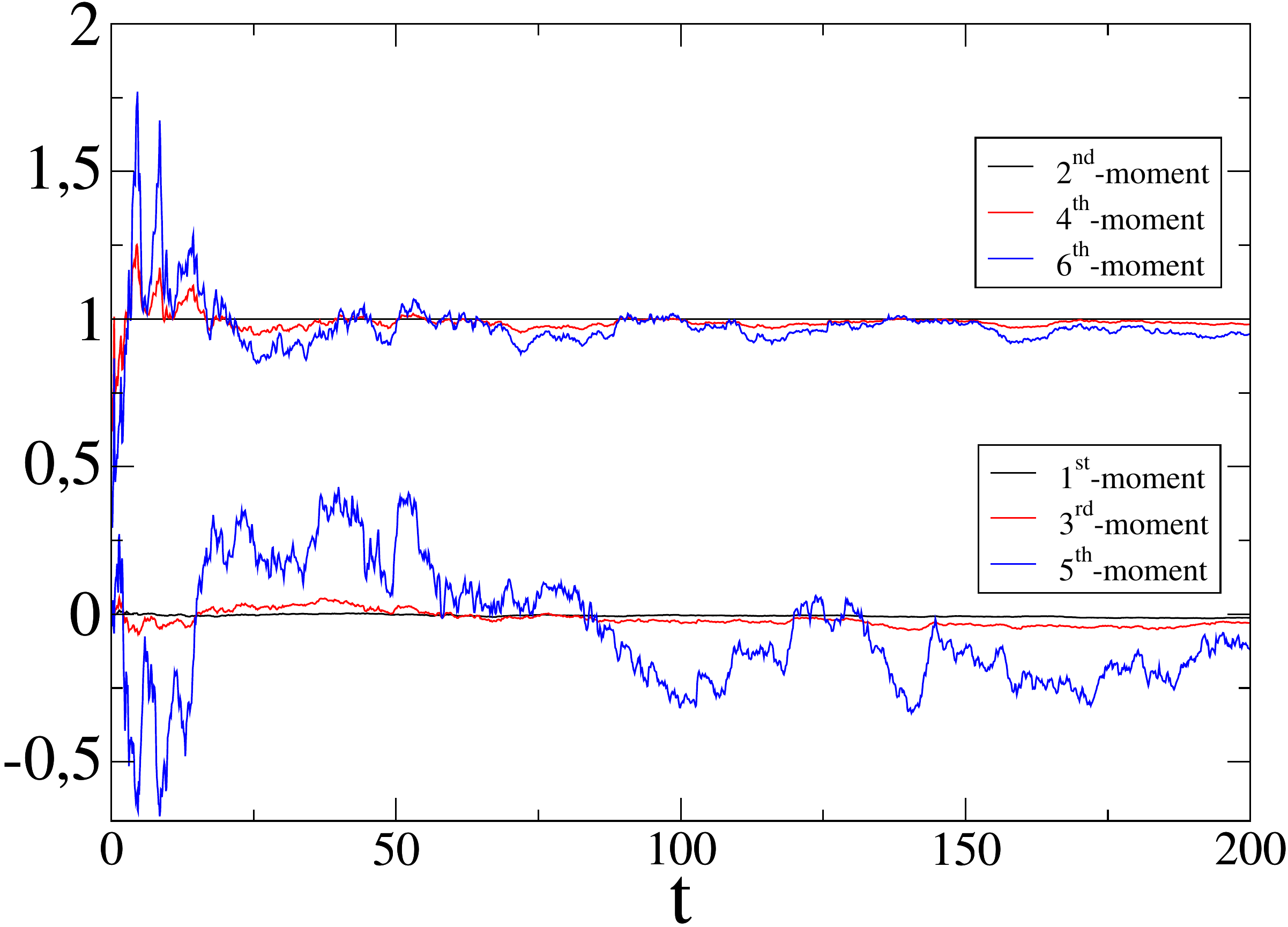}
\caption{Moments of $P_j(t)$ rescaled as (\ref{Podd})-(\ref{Peven}). 
System size $N = 10^6$, sample size $N_{\mathrm{s}} = 10^5$,
total simulation time $t = 200$, time step $\delta t = 0.1$.  }
\label{moments}
\end{figure}

To further check whether $P_j$ is a brownian motion, we can analyze the moments of its distribution to see whether they satisfy \cite{Ok10,KlPlSc03}
\begin{eqnarray}
	\frac{\left\langle P_j^k(t)\right\rangle}{S^k(t)} = 0, 
	\qquad\qquad \text{for odd $k$} 
	\label{Podd} \\
	\frac{\left\langle P_j^k(t)\right\rangle}{S^k(t) \cdot 1 \cdot 3 \cdot\cdot\cdot (k - 1)} = 1, 
	\qquad \text{for even $k$.}
	\label{Peven}
\end{eqnarray}
We run this test with the same configuration used in figure \ref{squared} and plot the results in figure \ref{moments}.
Both figures show that numeric simulation is in good agreement with the analytical results of the previous section. The strong fluctuations around the expected values in figure \ref{moments} are associated with the finite number of particles used in the simulation.

\section{Conclusions}

Starting from a ``particles in monokinetic beams"\ initial condition (illustrated by figure \ref{inibeam}) approximating a waterbag, 
we show analytically that the velocities of particles in the repulsive XY HMF $N$-body system display 
Brownian corrections to the ballistic motion implied by the Vlasov limit for $N \to \infty$,
and that these corrections propagate initial independence (molecular chaos). 

As we show in equation \eqref{eomninf}, the motion of particles in velocity space can be written as a stochastic differential equation driven by Wiener processes which are due to particle interaction, 
\textrm{i.e.} the coupling of the particles with the mean field quantity of the system. 
In such systems, the mean field forces are usually postulated as white noises based on numerical evidence 
(see \cite{EtFi11} and references therein) ; 
here, we show rigorously that process \eqref{resmag} properly rescaled as \eqref{3W} converges to a Wiener process, 
and moreover that particles in the same mean field do behave independently of each other 
(thanks to their different initial positions and velocities).
We present a lower time estimate for the validity of our approximations along with numerical results confirming our findings.

\section{Acknowledgements}

\quad This work benefited from fruitful discussions with T.~M. da Rocha Filho 
and participants to the XIV Latin American Workshop on Nonlinear Phenomena. 
Comments by D.~D.~A. Santos are gratefully acknowledged, as well as constructive comments by the anonymous referees. 
Author B.~V. Ribeiro acknowledges CAPES for financial support.


%
\end{document}